# Coherent Oscillations in an Exciton-Polariton Josephson Junction

K. G. Lagoudakis[1], B. Pietka[1], M. Wouters[2], R. André[3], B. Deveaud-Plédran[1]

[1]ICMP, Ecole Polytechnique Fédérale de Lausanne(EPFL),1015 Lausanne, Switzerland.
[2]ITP, Ecole Polytechnique Fédérale de Lausanne(EPFL),1015 Lausanne, Switzerland.
[3]Institut Néel, CNRS, Grenoble, France.

**Abstract:**
*We report on the observation of spontaneous coherent oscillations in a microcavity polariton bosonic Josephson junction. The condensation of exciton polaritons takes place under incoherent excitation in a disordered environment, where double potential wells tend to appear in the disordered landscape. Coherent oscillations set on at an excitation power well above the condensation threshold. The time resolved population and phase dynamics reveal the analogy with the AC Josephson effect. We have introduced a theoretical two-mode model to describe the observed effects, which allows us to explain how the different realizations of the pulsed experiment have a similar phase relation.*

One of the most striking manifestations of the quantum collective behavior of matter is the Josephson effect: it allows for a classically forbidden current to flow without dissipation [1]. Traditional Josephson junctions are built by means of superconductors separated by thin non conducting oxide barriers and they have drawn an intense interest for their striking phenomenology. A bosonic analog is the so called Bosonic Josephson junction (BJJ) where two macroscopic populations of bosons are trapped in a double well geometry. Bose Einstein condensation of dilute atomic gases has allowed for the realization of BJJs by trapping two condensates in such double wells [2],[3]. Although oscillatory behavior of fluids is common in classical systems (i.e. an initially full water container connected to an empty one through a large pipe), the Josephson effect results in a dissipationless flow through classically forbidden barriers and produces an alternating current from a constant potential difference between the two wells. This phenomenon is directly related to one of the most notable properties of quantum fluids: macroscopic phase coherence [4]. The broken symmetry in such fluids allows for the description of all particles with a common one-body wavefunction and a common phase which in turn allows for a number of quantum effects to occur.

The possibility to engineer and dynamically tune atomic condensates, has allowed for all the possible regimes and their respective phenomenology to be observed. More specifically in the Josephson regime, the most remarkable phenomena are the AC and DC Josephson effects [5], the macroscopic quantum self trapping and the Josephson plasma oscillations [6].

The recent achievement of condensation of exciton polaritons in semiconductor microcavities [7] has provided a new field for the study of quantum phenomena related to the macroscopic phase coherence. Polaritons are Bosonic quasi-particles that occur in semiconductor microcavities as a result of the strong coupling between light (cavity photons) and matter excitations (quantum well excitons). Owing to their half-light nature, polaritons have an exceptionally low mass of the order of $10^{-5}$ the mass of free electrons, which allows for condensation at temperatures easily achievable by cryogenic means. This makes exciton polaritons very attractive in the sense that they are a high temperature quantum fluid. Having a limited lifetime of about 3psec, they leak out of the microcavity and thus have to be continuously replenished, a process that renders polariton condensates out of equilibrium. It is because of this aspect that polariton condensates provide a very rich phenomenology [8]. Another advantage of the system under consideration is that all the condensed polariton properties are imprinted to the luminescence emitted by the condensate so one can gain access to both phase and population by optical imaging and interferometric methods.

Here we have set out to investigate Josephson effects in a BJJ formed by two spatially separated condensates of exciton polaritons in a CdTe microcavity. This work clearly demonstrates AC Josephson oscillations evidenced through relative population and relative phase oscillations between the two trapped condensates. This observation implies the existence of a synchronization mechanism between different experimental realizations which we understand in depth with the two mode model that we have developed to describe the observed phenomenology.

The condensate pair is trapped in a naturally occurring disorder double potential well which is set during the sample growth. These structures tend to appear more frequently in linear disorder valleys, see Fig. 1(a). The sample employed here is the same as the one used in our previous studies [9]. We employed pulsed non resonant excitation in conjunction with a streak camera at the detection that provided the necessary temporal resolution. The population oscillations were revealed by forming the image of the real space energy integrated photoluminescence (Fig. 1a) on the entrance slit of the streak camera. The BJJ that we studied here, consisted of two wells of about $2\,\mu m$ size each, separated by a barrier of approximately $0.5\,\mu m$. From the spectroscopic data we estimated the depth of the confining potential to be approx. 5.6meV with respect to the 2D polariton average energy. Remarkably, the Josephson oscillations only show up above a threshold excitation power $P_{OT}$ that is higher than the power required for condensation $P_{CT}$. At powers $P_{CT} < P < P_{OT}$, coherence between the two wells is observed, but the dynamics of the system do not show any interesting features and no population exchange is seen between the two trapped condensates as clearly shown in Fig. 1(b).

On the contrary when increasing the excitation power above the oscillation threshold then the behavior changes dramatically and one clearly sees an exchange of particles between the two wells Fig. 1(c). Additionally, to assess that the polariton currents are driven by the phase difference between the two wells, one needs the relative phase difference $\Delta\theta(t) = \theta_L(t) - \theta_R(t)$ with $\theta_{L,R}$ the absolute condensate phase in the respective wells. This is accessible only by means of interferometric measurements and for this purpose we used the modified Michelson interferometer in a mirror-retroreflector configuration with active stabilization [7] to interfere the luminescence of the left well with that of the right-one. The density of fringes in the interferograms is set by the displacement of the retroreflector which applies an effective angle of the reflected beam when passing through the imaging lens thus giving an interference term of the form $\cos\left[k_I \cdot x + \Delta\theta(t)\right]$, $\Delta\theta(t)$ being the relative phase difference. One can directly see that, when the two condensates have a constant phase difference, the fringes are straight in time (Fig. 2b) whereas, if their phase increases linearly in time, the fringes will appear tilted (Fig. 2c). By scanning the relative phase of the two interferometer arms over $3\pi$ we were able to fit the function $A(x,t) \cdot \cos\left[\phi + \Delta\theta(x,t)\right]$ at each pixel $(x,t)$ and thus retrieve the residual phase $\Delta\theta(x,t)$ up to the unknown constant phase $\phi$ induced by the initial position of the interferometer.

The extracted relative phase $\Delta\theta$ and normalized population difference $\Delta N = (N_L - N_R)/(N_L + N_R)$ are plotted in Fig. 3. Density and phase oscillations show the expected relation: the measurements are consistent with a Josephson current that flows from the well with the lower phase to the one with higher phase and vanishes when the phase difference is zero. We thus observe density oscillations between two macroscopically occupied polariton states driven by their phase difference, i.e. the bosonic Josephson effect

[10]. A closer look in the population dynamics reveals that there is a strong imbalance of the population at the first moments of the oscillations (between 35 and 65 psec) which is due to a higher energy level that does not contribute to the beatings and dies out quickly within the same temporal window. Time integrated spectrally resolved studies (not shown) allow to observe the emission lines of the two beating states, their energy difference corresponding to the beating frequency, and the slightly higher energy state that has no oscillatory signature in the temporal response.

The relative phase between the two wells is expected to increase linearly in time. Modulo $2\pi$, this should in turn give a perfect saw tooth form of the relative phase evolution from $-\pi$ to $\pi$. Contrary to this, we observe a smoothed saw tooth phase evolution with much smaller amplitude ($-0.3\pi$ to $0.3\pi$).

It is actually not trivial that Josephson oscillations can be observed under the present conditions. Indeed, the pulsed experiment is repeated millions of times and starts every time with an arbitrary relative phase between the condensates in the two disorder minima. One would therefore rather expect to see oscillations in a single run of the experiment, but not in our case of realization averaged signal.

In order to gain more insight in the origin and nature of Josephson oscillations that we observe, we compare our observations with the predictions of a two polariton mode that takes into account the polariton losses and their replenishment from the exciton reservoir [9],[11],[12]:

$$i\frac{d}{dt}\psi_{L,R} = \left\{\varepsilon_{L,R} + g_{L,R}|\psi_{L,R}|^2 + \frac{i}{2}\left[R(n_{R2;L,R}) - \gamma\right]\right\}\psi_{L,R} + J\psi_{R,L} + \xi_{L,R} \qquad (1)$$

Here $\varepsilon_{L,R}$ are the energies of the modes in the absence of the tunneling coupling $J$ and $\gamma$ is the linewidth in the linear regime. The nonresonant excitation enters the polariton dynamics through the stimulated relaxation rate $R(n_{R2})$. The exciton reservoir density $n_{R2}$ consists of the active excitons that fulfill energy and momentum conservation for scattering into the lower polariton branch. Their density is modeled with the rate equation

$$\frac{dn_{R2;L,R}}{dt} = \gamma_{R12}\left(\frac{n_{R1;L,R}}{r_{12}} - n_{R2;L,R}\right) - \gamma_{R2}n_{R2;L,R} - \alpha R(n_{R2;L,R})|\psi_{L,R}|^2 \qquad (2)$$

Here $r_{12}$ is the steady state ratio between inactive and active excitons. The active exciton density is normalized to be unity at the condensation threshold and $\alpha$ is the depletion factor of this reservoir due to the scattering into the lower polariton branch. The inactive exciton density $n_{R1}$ dynamics is given by

$$\frac{dn_{R1;L,R}}{dt} = \gamma_{R12}\left(\frac{n_{R1;L,R}}{r_{12}} - n_{R2;L,R}\right) - \gamma_{R1}n_{R1;L,R} + P_{L,R} \qquad (3)$$

The noise term $\xi$ models the spontaneous decay and scattering events, and can be explicitly computed within the truncated Wigner approximation [11]. The most important stochastic element is however the initial condition of the polariton field: it starts in a vacuum state with random phase.

Fig. 4 shows two simulations of the dynamics under pulsed excitation, both above the threshold for condensation and, interestingly, the Josephson oscillations survive the realization averaging only for the higher excitation power (right hand side panels). Our simulations thus reproduce the experimental observation that the threshold for Josephson oscillations $P_{OT}$ is higher than the one for condensation $P_{CT}$. Fig. 4 (a,b) shows multiple realizations of the density/phase evolution for $P_{CT} < P < P_{OT}$. Individual realizations of the density-phase evolution clearly show oscillations. In the average over many runs of the experiment (Fig. 4c) these oscillations have however disappeared, because the different realizations are not in phase. Only at higher power (see Fig. 4(d,e,f)), do the oscillations share some common phase and survive the realization averaging.

The physical reason is the following. At high excitation intensity, the growth rate is much larger than the tunneling between the two wells $R \gg J$ and the Josephson currents can be neglected during the exponential growth stage of the condensate formation. This implies that the initial randomness of the relative phase does not affect the density dynamics that is therefore deterministic up to small fluctuations. The dynamics of the relative phase $\Delta\theta(t) = \theta_L(t) - \theta_R(t)$ is governed by

$$\frac{d\Delta\theta}{dt} = J\left(\sqrt{\frac{N_L}{N_R}} - \sqrt{\frac{N_R}{N_L}}\right)\cos(\Delta\theta) + \varepsilon_R - \varepsilon_L + g_R N_R - g_L N_L, \qquad (4)$$

Where we have introduced the density and phase of the fields $\psi_j = \sqrt{N_j}e^{i\theta_j}$, $j = L,R$. The emergence of a fixed relative phase shortly after the condensate formation can be easier understood for $\varepsilon_R = \varepsilon_L$. Under asymmetric pumping, one of the two densities is much larger than the other one. Eq. (4) then implies that the relative phase will be quickly driven to $\Delta\theta = \pi/2$, the zero of the cosine. Our simulations show that also when $\varepsilon_R \neq \varepsilon_L$, a deterministic relative phase is reached in the early stage of the dynamics, irrespective of the initial condition (see Fig. 4d).

A second question we can address with our model is whether our observations are rather related to the Josephson plasma oscillations or the AC Josephson effect. The latter effect is driven by the energy difference between the two potential minima, where the former originates from an initial condition that is different from the steady state. A previous study [13] of plasma oscillations in a nonequilibrium polariton condensate under continuous wave excitation has shown that plasma oscillations feature a damped sinusoidal dynamics, in contrast to the persistent saw-tooth like behavior that is observed in our experiments. The polariton analog of the AC Josephson effect on the other hand occurs when the energy difference between the two wells is too big for a synchronized solution to exist [14],[15]. The beatings between the two modes are responsible for persistent density oscillations and a saw-tooth phase profile, as we observe in our experiments. The oscillations between two wells with different frequency condensates are entirely analogous to the AC Josephson oscillations where the difference in chemical potential is imposed by an external voltage.

An unexpected feature of the saw tooth oscillations in Fig. 3 is that their amplitude is much below the expected $2\pi$. This property of the realization averaged phase can be traced back to the fact that the phase evolution is steeper for large $\Delta\theta$ (see Fig. 4d). As a consequence large values of $\Delta\theta$ do not contribute much to the average phase difference. The physical reason is that large phase differences go together with a low population in one of the two wells. When one of the populations becomes very small, the Josephson current quickly reverts so to replenish it. A further consequence of the rapid phase variation is that the shot to shot variations are more significant at those times. This results in a reduced coherence, seen both in the experiments and theoretical simulations (not shown in the figures).

Finally, our two-mode model also allows us to understand the role of the polariton-polariton interactions on the Josephson oscillations. In the absence of interactions, the oscillation frequency is set by the detuning and tunneling coupling. When the blue shift due to polariton-polariton interactions $\mu_j = gN_j$ becomes comparable to the detuning $\Delta\varepsilon$, interactions make the oscillation frequency increase for increasing pump intensity. In experiments however, no such behavior is observed. Contrary to this, a slight decrease of the frequency is observed. We therefore conclude that polariton-polariton interactions have a negligible effect on the oscillation dynamics and that our polaritonic BJJ is in the so-called Rabi regime [17]. The slight decrease in oscillation frequency can be reproduced in the theory when taking into account the interactions between the polaritons and the exciton reservoir. If the low energy well is pumped with higher intensity, the blue shift due to the excitons will shift it up more than the other one and therefore the oscillation frequency decreases.

In this work we have provided experimental evidence of coherent oscillations in a Bosonic Josephson junction formed by two exciton polariton condensates created under nonresonant pulsed excitation in a double well geometry. We have shown the simultaneous existence of population and phase oscillations and have carefully evaluated their amplitudes and average values. The saw-tooth like oscillations persist for the whole life time of the condensates and are the polariton analog of the AC Josephson regime. Our theoretical analysis shows that the Josephson oscillations survive the realization averaging linked with the random initial phase, though with reduced amplitude, thanks to a spontaneous synchronization mechanism that only appears above a critical pump intensity. The present realization of the Josephson effect relies on the accidental presence of coupled wells in our disordered microcavity. Controlled growth of polariton traps with mesa structures [18] and micropilars [19] will allow for a more systematic analysis of the different regimes of Josephson oscillations, where polariton-polariton interactions can play a more pronounced role. Also, combining nonresonant with resonant excitation will provide full control over the initial conditions of the Josephson oscillations.

**ACKNOWLEDGEMENTS**
The authors wish to thank Le Si Dang, Vincenzo Savona, Yoan Leger, Iacopo Carusotto, Davide Sarchi, Maxime Richard and Augustin Baas for continuous stimulating discussions. The work was supported by the Swiss National Research Foundation through NCCR "Quantum Photonics".

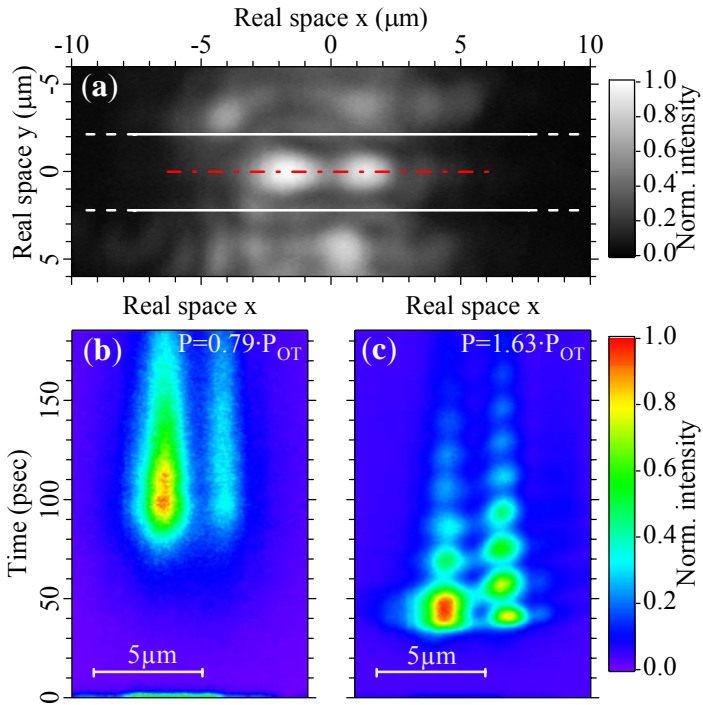

FIG. 1 (Color online). (a) Time integrated real space image of the luminescence. The dash-dotted line is the line selected by the streak camera and the solid lines with dashed edges enclose the linear disorder valley in which the double well potentials are frequent. Here we have chosen the center of the axes as the center of the polariton Josephson junction. (b) Time resolved photoluminescence along the dash-dotted line in (a) for excitation power below the oscillation threshold. (c) Same as in (b) but for an excitation power above the oscillation threshold. One can easily see the striking difference between the case with no population oscillations and the case where oscillatory exchange is strongly pronounced. The incident pulse defines the zero time.

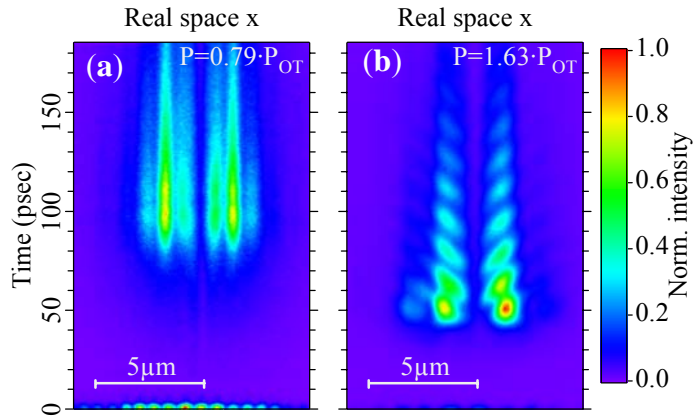

FIG. 2 (Color online). Time resolved interferograms along the dash-dotted line of Fig. 1(a) formed by the superposition of the luminescence of each well with its neighboring. (a) Excitation power below the oscillation threshold. The lack of oscillatory behavior is clearly seen with the constant phase relation between the two wells leading to straight fringes in time. (b) Same as in (a) but for an excitation power above the oscillation threshold. The oscillatory current of particles is due to the linearly increasing relative phase in time which provokes the tilting of the fringes.

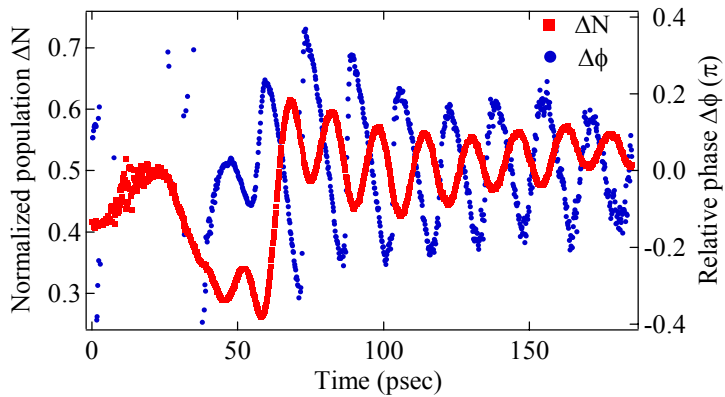

FIG. 3 (Color online). Temporal evolution of relative well population and relative well phase. The relative population (squares) oscillates symmetrically around zero and the phase (circles) is following the population oscillations. A striking feature is that the phase is not oscillating between $-\pi$ and $\pi$ in a saw-tooth manner as expected but rather shows small amplitude saw-tooth like oscillations. The region between 5 and 35 psec had a very low overall intensity and thus the relative phase could not be calculated.

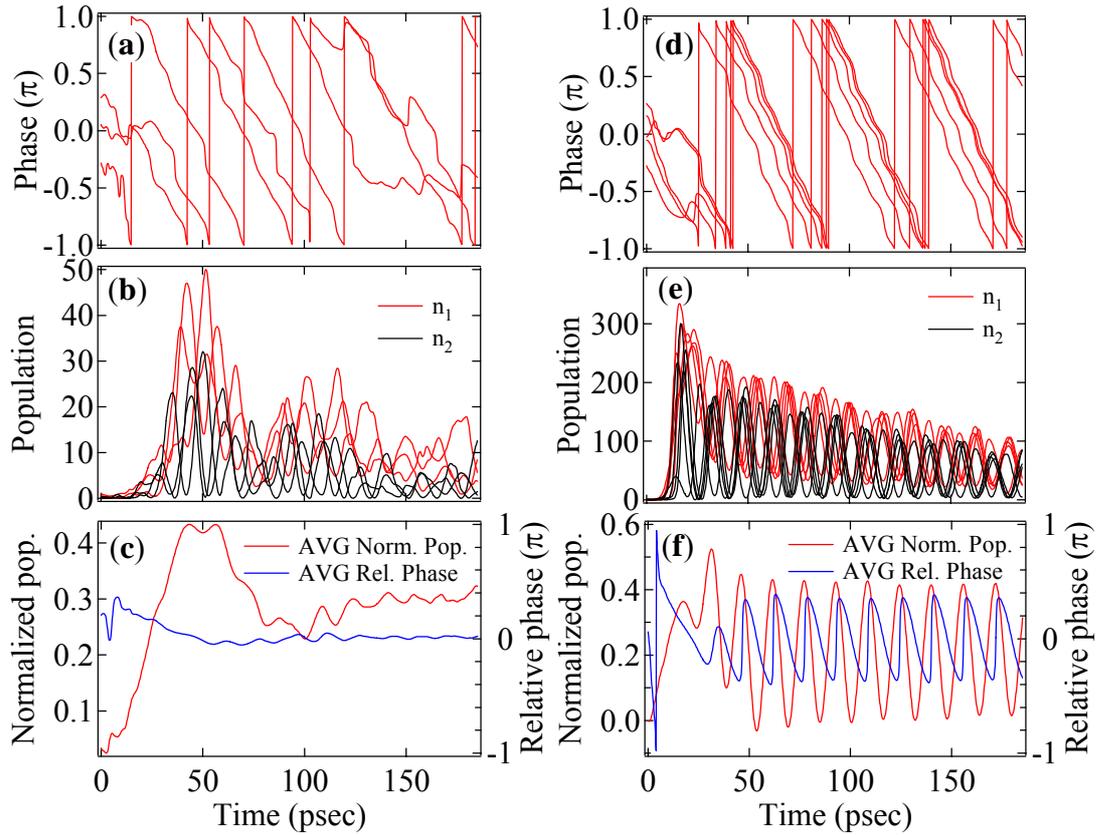

FIG. 4 (Color online). Theoretical simulations: (a) Phase evolution for individual realizations for low excitation powers. (b) Populations in left (red) and right (black) well for the three realizations of (a). (c) Averaged relative phase modulo $2\pi$ and normalized populations from (a), (b). (d), (e) and (f) are the same as the panels (a), (b) and (c) only this time for excitation intensity above the oscillation threshold. Both the phase and the populations have very similar behaviors for different realizations demonstrating the effect of the phase locking mechanism. The non negligible fluctuations of the phase are the reason of the smoothed saw tooth phase profile [16].